# Broadband plasmonic nanoantennas for multi-color nanoscale dynamics in living cells

*Maria Sanz-Paz[†,¥,#], Thomas S. van Zanten[†,‡,#], Carlo Manzo[†,£], Mathieu Mivelle[§,*], Maria F. Garcia-Parajo[†,⊥,*]*

[†] ICFO-Institut de Ciencies Fotoniques, The Barcelona Institute for Science and Technology, 08860 Barcelona, Spain

[¥] Department of Physics, University of Fribourg, Chemin du Musée 3, Fribourg CH-1700, Switzerland;

[‡] National Centre for Biological Sciences, Bangalore, India

[£] Facultat de Ciéncies, Tecnologia i Enginyeries, Universitat de Vic – Universitat Central de Catalunya, C. de la Laura 13, 08500 Vic, Spain

[§] Sorbonne Université, CNRS, Institut des NanoSciences de Paris, UMR 7588, 75005 Paris, France

[⊥] ICREA, Pg. Lluis Companys 23, 08010 Barcelona, Spain

[#] Equally contributing authors

**Corresponding Authors**

*Email: maria.garcia-parajo@icfo.eu

*Email: mathieu.mivelle@sorbonne-universite.fr

**ABSTRACT**

Recently, the implementation of plasmonic nanoantennas has opened new possibilities to investigate the nanoscale dynamics of individual biomolecules in living cell. However, studies have yet been restricted to single molecular species as the narrow wavelength resonance of gold-based nanostructures precludes the simultaneous interrogation of different fluorescently labeled molecules. Here we exploited broadband aluminum-based nanoantennas carved at the apex of near-field probes to resolve nanoscale-dynamic molecular interactions on intact living cell membranes. Through multicolor excitation, we simultaneously recorded fluorescence fluctuations of dual-color labeled transmembrane receptors known to form nanoclusters in living cells. Fluorescence cross-correlation studies revealed transient interactions between individual receptors in regions of ~60 nm. Moreover, the high signal-to-background ratio provided by the antenna illumination allowed us to directly detect fluorescent bursts arising from the passage of individual receptors underneath the antenna. Remarkably, by reducing the illumination volume below the characteristic receptor nanocluster sizes, we resolved molecular diffusion within nanoclusters and distinguished it from nanocluster diffusion. Spatiotemporal characterization of transient interactions between molecules is crucial to understand how they communicate with each other to regulate cell function. Our work demonstrates the potential of broadband photonic antennas to study multi-molecular events and interactions in living cell membranes with unprecedented spatiotemporal resolution.

In recent years, the compartmentalization of biomolecules in space and time has emerged as a primary mechanism that regulates cellular function[1–5]. At the plasma membrane level, extensive research has demonstrated that multiple molecules, such as proteins and lipids, interact in a dynamic fashion creating transient nanoscale compartments of functional activity[6–10]. Such findings were revealed owing to the recent development of different optical techniques aimed at improving both spatial and temporal resolution beyond that of conventional diffraction-limited optical methods[10–15]. Yet, monitoring dynamic multi-molecular interactions in living cells at the nanoscale remains challenging.

With the advancement of single-molecule and super-resolution approaches, multiple techniques have been implemented aiming to reduce the illumination volume set by diffraction, thus enabling single-molecule dynamic studies at high labeling conditions in living cells on the nanoscale. For instance, stimulated emission depletion microscopy (STED)[16] and metallic nanoapertures[17–19] reduce the illumination area down to 50-200 nm in diameter. However, in the case of STED, both the high laser powers required and the increased photobleaching constitute significant drawbacks for its routine application in living cells[20]. Moreover, although dual-color STED is nowadays widely used for super-resolution imaging in fixed cells[21], its extension to living cells for multi-molecular dynamic studies using fluorescence cross-correlation spectroscopy (FCCS) remains highly challenging. In the case of subwavelength apertures, there is a compromise between volume confinement and light throughput since the effective power density decays as the fourth power of the aperture size. This severely limits their practical use in nanoscale studies, since aperture dimensions must be kept around 150-200 nm to provide sufficient excitation power. Nanoapertures and zero-mode waveguides (ZMW) have been used to perform simultaneous two-color FCCS in solution[22], on supported lipid bilayers[18] and live cells[18]. However, they suffer from the same throughput drawbacks as the single-color approaches.

Photonic antennas take advantage of electromagnetic resonances to enhance the optical field at nanometric dimensions[23], reducing the observation volume to a few zeptoliters and thus enabling the detection of single molecules in highly concentrated solutions[23,24]. These exciting results have prompted the search for different antenna nanofabrication strategies and geometries for nanoscale studies under biologically-relevant scenarios[25–31]. For instance, using nanoantenna arrays, we have been able to follow lipid diffusion in both model membranes and living cells[32–34]. Importantly, such measurements proved the existence of transient nanodomains in the membrane of living cells as small as 10 nm in size[32], demonstrating the great potential of photonic antennas to measure molecular diffusion and unravel nanoscale heterogeneities in intact living cell membranes.

Nevertheless, despite their advantages, several challenges still limit the broad application of these devices for biological membrane studies. Because of the strong field gradients of the antenna near-field, the antenna needs to be positioned close to the fluorescence molecules (~10nm) such that the largest enhancement and spatial confinement are reached. This is commonly achieved by preparing lipid bilayers or seeding the cells on top of the antenna substrate, requiring careful sample preparation to minimize unwanted interactions between the sample and the underlying substrate that can potentially affect the diffusion of the molecules. In addition, most antenna designs are only resonant in a narrow wavelength range, restricting experiments to a single color. Here, we address these two limitations by implementing self-standing broadband photonic antennas fabricated at the apex of near-field scanning probes. We show that, by maintaining the antenna stationary within 10 nm above intact living cell membranes, lipid diffusion can be recorded on regions of ~50 nm in size. We further demonstrate the capability of broadband antennas for FCCS at the nanoscale on living cells. Using this approach we resolve receptor interactions within nanoclusters and, importantly, discriminate molecular diffusion within nanoclusters from nanocluster diffusion in the plasma membrane of living cells.

**RESULTS AND DISCUSSION**

In our experiments, photonic antennas were fabricated at the apex of tapered near-field probes and relied on a near-field scanning optical microscope (NSOM) for 3D positioning control of the antenna over the living cell membrane (Figure 1A). Laser light ($\lambda = 488$ nm and/or $\lambda = 633$ nm) was coupled to the back end of the near-field probe and guided toward the antenna. The near-field light exiting the antenna excited the sample that was maintained stationary with respect to the antenna. Fluorescence intensity fluctuations arising from the passage of molecules diffusing through the antenna illumination volume were collected using a high-NA objective (NA=1.3), filtered out from the excitation light, and sent to two single-photon counting avalanche photodiodes (APD) arranged for dual color spectral detection. Two photon-counting units were used to record the fluorescent photon arrival times, which were then processed by a software correlator[35]. For antenna design, we chose bowtie nanoaperture antennas (BNAs) carved on aluminum-coated optical fibers using a focused ion beam (FIB)[36]. Our fabrication approach allows for extreme reproducibility of BNA probes, with a gap between the metallic arms of ~50 nm[36]. Moreover, BNAs provide optical throughputs of ~$10^{-3}$, three orders of magnitude larger than circular nanoaperture probes of similar dimensions[36].

For accurate position control of the antenna over the cell membrane surface, we relied on a shear-force feedback loop with high sensitivity under liquid conditions[37]. The feedback loop maintained the antenna-sample distance at ~10 nm with an error of ±1 nm under liquid conditions (Figure S1). This approach has two advantages as compared to antennas fabricated on substrates. First, because of the controlled distance separation, it minimizes unwanted sticky interactions between the antenna and the membrane that might alter the diffusion of molecules. Second, diffusing fluorescent molecules experience the same degree of near-field enhancement and confinement by the antenna, as opposed to antenna substrates in which distance variations might occur due to sample preparation imperfections and/or axial membrane fluctuations. Indeed, we occasionally membrane movements as high as 25 nm over 10 s (Figure S1) that our feedback loop was able to track and compensate for with an accuracy of ±1 nm. This shows that our overall approach maintains a constant axial distance between the membrane and the antenna despite potential membrane fluctuations during the measurements.

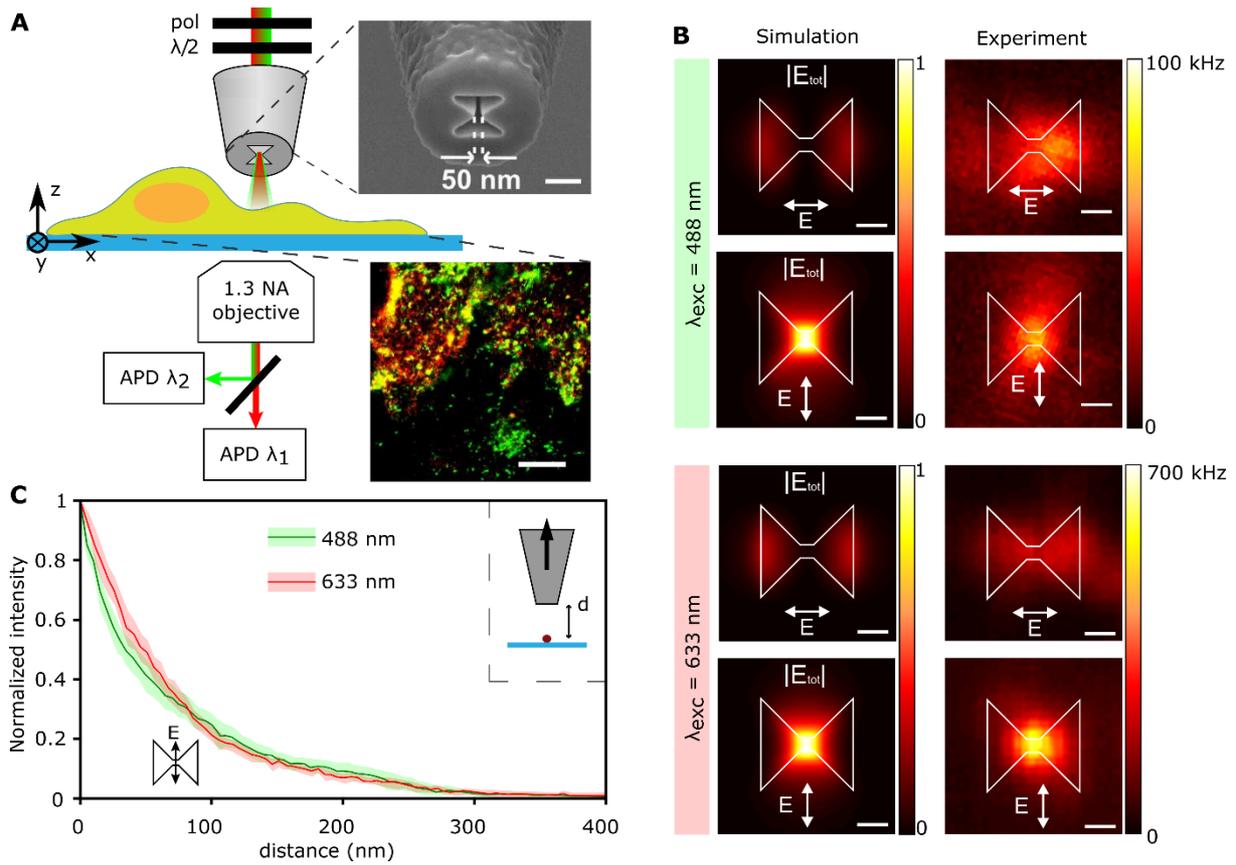

**Figure 1: Operation and assessment of the optical performance of a photonic antenna probe to measure molecular diffusion on living cell membranes.** **(A)** Schematics of the experimental setup. The BNA is engineered at the apex of an NSOM probe whose 3D position with respect to the sample is controlled with nanometer precision. The antenna is kept stationary with respect to the cell membrane and illuminates a nanoscopic area. Fluorescence fluctuations from diffusing molecules are collected via the objective and sent to detectors for multicolor detection. The top inset shows a representative SEM image of a BNA probe (scale bar: 100 nm). The bottom inset displays, as an example, a dual-color confocal image of a living CHO cell attached to the substrate, with surface receptors labeled with two different fluorophores (scale bar: 5 μm). **(B)** FDTD simulations of the total electric field at λ = 488 nm (upper panel set) and λ = 633 nm (lower panel set) excitation together with the experimentally obtained fluorescence intensity maps from two spectrally different 20 nm beads excited by a BNA under the two main orthogonal excitations. For both wavelengths, the field is maximally confined and enhanced for excitation polarization transversal to the BNA gap. Scale bar: 100 nm **(C)** Experimentally measured fluorescence intensity from 20 nm beads excited by a 50 nm gap BNA antenna probe as a function of the axial probe-sample distance under different wavelength excitations and for a polarization transversal to the BNA. The solid lines represent the mean, and the shadowed areas correspond to the standard deviation from multiple retraction curves.

To assess the spectral response of the BNAs, we performed FDTD simulations for different antenna gap sizes. Consistent with our earlier simulations[36,38], the response of the BNA is broadband over the whole range of the visible spectrum regardless of the gap size (Figure S2), showing its potential for multicolor excitation with comparable enhancement and confinement, opening the possibility for FCCS experiments at the nanoscale.

We further performed FDTD simulations to determine the (*x,y*) near-field intensity distributions of the BNAs for different wavelength excitations and optical fields transversally or longitudinally polarized with respect to the BNA gap (Figure 1B). The simulations assume a BNA size of 300x300 $nm^2$ and a gap size of 30 nm. For both ends of the visible spectrum ($\lambda$ = 488 and $\lambda$ = 633 nm), the field is highly confined and enhanced in the gap region of the BNA for a transversally polarized optical field (Figure 1B). However, for longitudinally polarized excitation and regardless of the wavelength used, the resonance is lost, and the field spatially delocalizes away from the gap (Figure 1B). To experimentally validate the simulations, we imaged 20 nm beads embedded in a thin polymer layer using a 50 nm gap BNA probe for the two excitation polarizations and wavelengths. In agreement with our simulations, (*x,y*) fluorescence distributions obtained from individual beads showed a larger enhancement and confinement for transversally polarized excitation regardless of the wavelength used (Figure 1B), confirming the broadband resonant character of the BNA. In addition, the field enhancement at both excitation wavelengths is directly estimated by calculating the ratio of the field detected for transversal and longitudinal polarizations, resulting in a 3.3-fold increase for $\lambda$ = 633 nm and a 1.7-fold increase for $\lambda$ = 488 nm. This larger enhancement measured at $\lambda$ = 633 nm agrees well with the expected spectral response of our BNA, which is about two-fold higher at $\lambda$ = 633 nm as compared to $\lambda$ = 488 nm (see Figure S2, for a 50 nm gap).

To evaluate the degree of confinement of the electric field in the axial direction for the two excitation wavelengths, the fluorescent intensity of individual beads *vs.* the antenna-sample axial distance was recorded (Figure 1C and Figure S3). A single exponential fitting of the transversely excited BNA (Figure 1C) yields an axial field penetration (1/e) of (66.6 ± 0.6) nm at $\lambda$ = 633 nm and (72.0 ± 1.0) nm at $\lambda$ = 488 nm. Notably, the standard deviation obtained from multiple approach-retraction experiments is very small, demonstrating the accurate axial control of the antenna position. Furthermore, taking the experimentally obtained confinement dimensions, we estimate the excitation volume of the BNA probe to be ~5·$10^5$ $nm^3$, more than two orders of magnitude smaller than the typical confocal volume. Overall, these results confirm the broadband response and nanoscale confinement of BNAs, highlighting their suitability for nanoscale multicolor experiments.

To first demonstrate the feasibility of BNA probes to study the lateral mobility components in the membrane of living cells, we used Atto647N conjugated phosphoethanolamine (PE) lipids [16,19]. CHO cells were allowed to spontaneously adhere to a glass coverslip for 48 hours and the fluorescent PE lipids were incorporated into the cell membrane (100-300 nM PE/BSA; two orders of magnitude higher labeling concentrations as compared to confocal) as described previously[16,19]. Fluorescence fluctuations from diffusing PE lipids were recorded using a BNA probe with a nominally gap size of 50 nm, excited at λ = 633 nm. For longitudinal BNA excitation, intensity bursts below 100 kHz (background ~30 kHz) were typically recorded (Figure 2A). In strong contrast, for transversal BNA excitation, intensity bursts of up to 600 kHz were detected (Figure 2B). Moreover, burst durations, i.e., passage time of the lipids through the antenna illumination area, were much shorter for transversal polarized excitation as compared to longitudinal. These two effects (increased fluorescence and shorter burst duration) confirm the polarization-dependent enhancement and spatial confinement provided by BNAs directly measured on living cells.

Fluorescence time traces of at least 5 seconds in length were autocorrelated for different experiments using $G(\tau) = \langle F(t) \cdot F(t+\tau) \rangle / \langle F(t) \rangle^2$, where $\tau$ is the delay (lag) time and $\langle \rangle$ indicates time averaging. Multiple autocorrelation functions (ACFs) were averaged and normalized for each excitation condition, i.e., confocal and the two BNA excitations (Figure 2C). A clear shift of the ACF curves towards shorter time lags was obtained when going from confocal to BNA transversal excitation. Since PE diffuses randomly within the membrane[19,32], the shortening in the diffusion times obtained upon BNA transversal excitation results from the reduced illumination area provided by the antenna gap.

To statistically confirm these results, we generated ACF curves from individual fluorescence traces and measured the amplitude of the ACF at $G(0)$ and the time at which the ACF decays to half of its amplitude, $\tau_{1/2}$. $G(0)$ inversely scales with the apparent number of fluorescing molecules in the illumination area $N$, while $\tau_{1/2}$ reports on the characteristic diffusion time of the lipids through the illumination area[16,19]. The distributions of $N$ and $\tau_{1/2}$ values over multiple ACFs are shown in Figure 2D. For a longitudinally excited BNA, the mean number of PE lipids and average transit time are $N = 8.5 \pm 1.8$ and $\tau_{1/2} = (12 \pm 3)$ ms, respectively. For transversal BNA excitation, the PE diffusion times become much shorter, with a mean average of $\tau_{1/2} = (1.8 \pm 0.5)$ ms (Figure 2D). Moreover, for randomly diffusing molecules such as PE[19,32], a reduction in the illumination area should scale linearly with a reduction in the number of molecules, $N$, as is indeed observed. Using the measured transit times we estimate a 12/1.8 = 6.7-fold reduction in the number of molecules. Nevertheless, the average number of molecules obtained for transversal excitation is $N = 4\pm2$ (Figure 2D), i.e., showing a reduction of only 2.1-fold with respect to longitudinal excitation. This apparent discrepancy can be well understood by the fact that, for transversal BNA excitation, the antenna is resonant, and the field is enhanced, leading to higher intensity emission from the fluorophores and, thus, an effective higher $N$. The field enhancement can thus be directly estimated from the ratio between the expected reduction of $N$ and the experimentally obtained values, yielding a 6.7/2.1 = 3.2-fold fluorescence enhancement. This value agrees excellently with the 3.3-fold enhancement obtained from measuring beads (Figure 1B), demonstrating that the antenna performance is fully maintained even in complex environments such as living cell membranes.

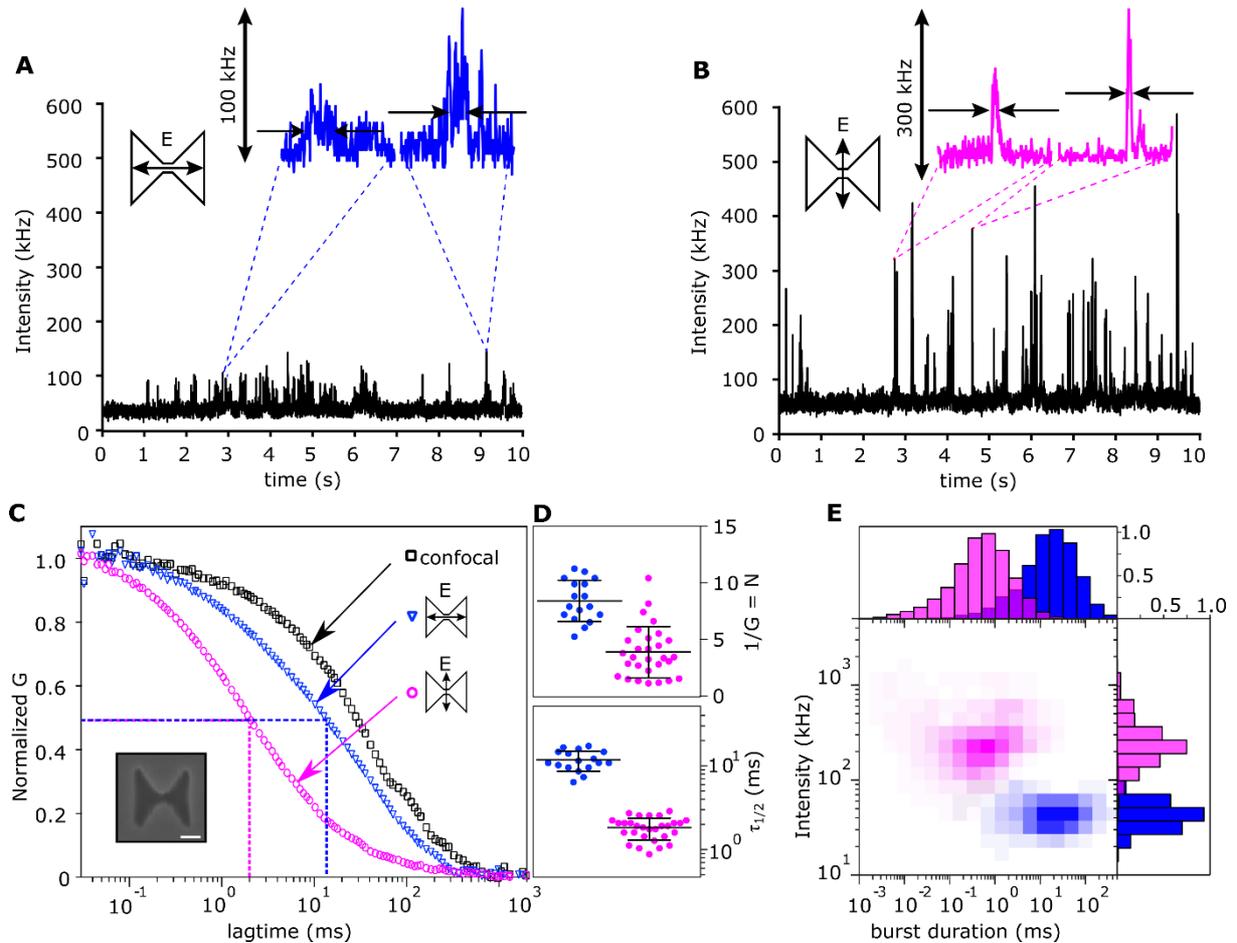

**Figure 2. Single lipid mobility in living cell membranes at the nanometer scale recorded with a BNA probe. (A, B)** Representative fluorescence time traces (1 ms bin) of Atto647N-conjugated PE diffusing in a living cell membrane for the two orthogonal excitation polarization conditions (shown in the insets) of the BNA at λ = 633 nm. Both traces were recorded at the same membrane position. The insets display 200 ms zoom-ins of different representative bursts with a 50 μs bin. Note that the overall constant background in both fluorescence time traces further indicates the fixed distance separation between the antenna and the cell membrane maintained by the feedback loop. **(C)** Normalized ACFs curves of PE-Atto647N diffusing on the cell membrane under confocal (black squares), longitudinally excited BNA (blue triangles), and transversally excited BNA gap (magenta circles) illumination. The dashed lines correspond to the $\tau_{1/2}$ values. **(D)** Distributions of the apparent number of molecules $N$ and characteristic diffusion times obtained from individual trace, **(E)** 2D plot together with population distributions of burst duration *vs.* burst brightness (average background corrected) directly extracted from multiple fluorescence traces for longitudinal (blue, over 1488 bursts) and transversal (magenta, over 3643 bursts) excitation to the BNA gap.

The large signal-to-background ratio afforded by BNA antennas allowed us to additionally perform burst analysis[32,39] over multiple fluorescence trajectories. Here, individual bursts likely correspond to the passage of single PE lipids transiting within the BNA illumination area.

We determined each burst's duration, i.e., transit time and background-subtracted intensity. Two discrete populations with different burst durations and intensities were unambiguously recovered for the two polarization excitation modes of the BNA (Figure 2E). Furthermore, a clear correlation between both parameters was obtained: shorter and brighter bursts for transversal BNA excitation, signatures of higher field confinement, and enhancement. Indeed, a ~30-fold shortening of the burst duration was obtained for transversal ($10^{-2}$ - $10^{1}$ ms, peak at 0.7 ms) *vs*. longitudinal ($10^{-1}$ - $10^{2}$ ms, peak at 20 ms) BNA excitations (Figure 2E). The peak values of the histograms are within the range of the $\tau_{1/2}$ values derived from the ACFs (Figure 2D), although the latter corresponds to average diffusion times over individual trajectories, whereas the burst duration histograms reveal the entire distribution of individually diffusing lipids. Furthermore, the effective confinement area provided by the BNA gap can be calculated directly from the recovered transient times, considering the reported diffusion coefficient of PE (0.5 μm$^2$/s)[16,19]. We obtain a confinement diameter (assuming a circular illumination profile) between 42 nm (taking 0.7 ms from the burst analysis) and 68 nm (taking 1.8 ms as derived from the ACF curves), which is within the range of the 50 nm gap size as measured by SEM. Finally, the burst brightness provided a direct measure of the increased excitation intensity given by the BNA. The background corrected burst intensity increased by 11-fold, from 100 kHz to 1100 kHz maximum value (Figure 2E), upon changing the BNA excitation from longitudinal to transversal, similar to earlier reports[36,40,41]. In summary, these results demonstrate the well-maintained optical performance of photonic antennas for studies in living cell membranes.

We further explored the potential of our broadband antennas for multi-color fluorescence autocorrelation spectroscopy (FCS) and FCCS studies at the nanoscale in living cells. Here, we focused on the transmembrane receptor DC-SIGN, a pathogen recognition receptor that forms nanoclusters ranging from 100-400 nm on both dendritic and CHO cells[42–47]. DC-SIGN nanoclustering has been reported to play a key role in the capture of a large variety of nanometric-sized viruses [43,45,46]. Although DC-SIGN nanoclustering has been studied using a broad range of techniques, including TEM[43,46], super-resolution microscopy[42], fluorescence recovery after photobleaching (FRAP)[47] and single-particle tracking (SPT)[43], it is not yet clear whether DC-SIGN nanoclusters are stable in time or assemble/disassemble transiently. Moreover, it is still an open question whether DC-SIGN receptors are mobile within nanoclusters or if the reported diffusions of DC-SIGN correspond to the mobility of the entire nanoclusters[43,47].

To address some of these questions, we labeled the DC-SIGN on a stably transfected on CHO cells at saturating conditions, using equimolar concentrations of Atto520- and Atto647N-conjugated to single-chain antibodies. Dual-color FCS and FCCS experiments at the nanoscale were performed using a BNA probe (50 nm gap size) simultaneously excited with λ = 488nm and λ = 633 nm (transversal polarization for both excitation wavelengths). Representative fluorescence traces recorded simultaneously in the two detection channels are shown in Figure 3A and Figure S4. Zoom-ins of coincident bursts indicate receptor co-diffusion within the same nanometric illumination area. Multiple single-color ACFs were normalized and averaged (Figure 3B). The shape of the ACF curves and the $\tau_{1/2}$ values (17±6 ms and 10±2 ms for Atto647N and Atto520, respectively) are comparable, confirming similar confinement areas for both excitation wavelengths. Taking into account the transit times and the excitation area estimated above from the ACF curves for λ = 633 nm excitation (3600 nm$^2$), we calculate an average diffusion coefficient for DC-SIGN of $D = A/(4\tau) = 5.3 \cdot 10^{-2}$ µm$^2$/s, which is within the range of our previously published values[43].

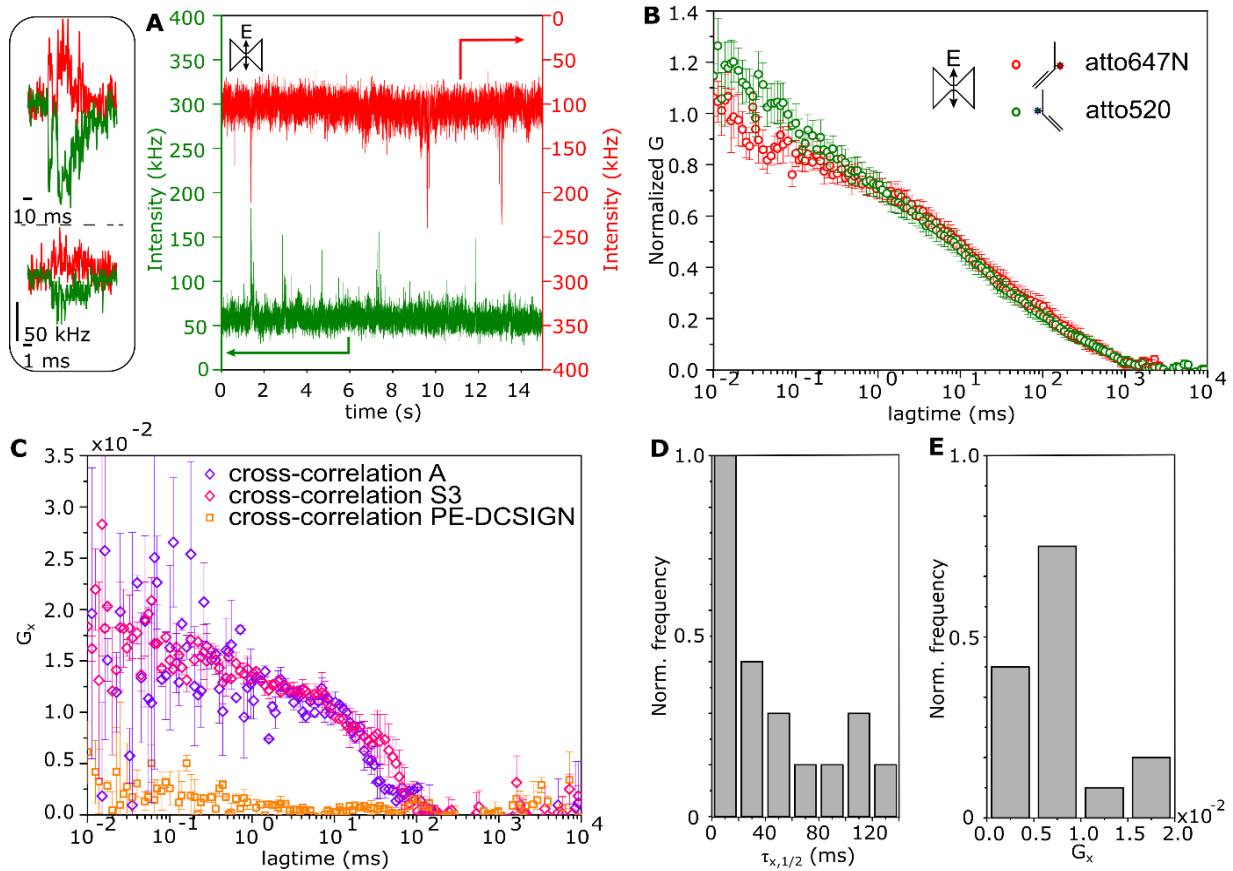

**Figure 3. Simultaneous dual-color detection of receptor diffusion and FCCS at the nanoscale. (A)** Representative fluorescent time trace (1 ms bin) of Atto520 (green) and Atto647N (red) conjugated to single chain antibodies and bound to DC-SIGN, expressed on CHO cells. Trajectories were generated using transversally polarized excitation of a BNA probe simultaneously excited with λ = 488nm and λ = 633 nm. Zoom-ins of some of the coincident bursts are shown in the insets. **(B)** Normalized ACFs corresponding to the fluorescent intensity traces of dual-labeled DC-SIGN. **(C)** CCFs of two different dual-color DC-SIGN traces (pink and purple diamonds). The orange curve corresponds to the cross-correlation of Atto520-DC-SIGN with Atto647N-PE, and is used as a negative control to show no specific co-diffusion. **(D, E)** Histograms of the co-diffusion times **(D)** and the cross-correlation amplitudes **(E)** were obtained from finding $\tau_{x,1/2}$ from the curves in Figure S5A.

We then generated cross-correlation functions (CCFs) (Figure 3C) for the traces shown in Figure 3A and Figure S4. Non-negligible cross-correlations amplitudes $G_x$ were retrieved for both traces, indicating co-diffusion of receptors in the same nanoscale volume and, thus, nanoscale interaction. In addition, average transit times of $\tau_{x,1/2}$ of 10.2±1.8 ms and 19±8 ms were obtained for the two different curves, indicating heterogeneity in DC-SIGN diffusion, in agreement with earlier SPT results[43]. As a control, we also recorded time traces of DC-SIGN labeled with Atto520 and the Atto647N-conjugated PE lipid analog. As expected, the CCF

13 of 23

curves are almost flat and remained close to zero, consistent with a lack of cross-correlation between DC-SIGN and the lipid diffusions, and validating our FCCS measurements at the nanoscale.

Multiple CCF curves of DC-SIGN obtained from different cells and/or membrane regions are shown in Figure S5A. The curves exhibit a broad distribution of cross-correlation times $\tau_{x,1/2}$ and cross-correlation amplitudes at $G_x(0)$ (Figure 3D, E). For comparison, the amplitude histogram of CCF curves from the traces with a lacking of cross-correlation is shown in Figure S5B. The broad range of characteristic $\tau_{x,1/2}$ times obtained from the CCF curves can arise from both correlated motions of individual DC-SIGN receptors within the same nanocluster or correlated motion between different nanoclusters. In either case, the broad distribution of $\tau_{x,1/2}$ indicates a range of interaction strengths of DC-SIGN receptors with their nano-environment that directly impinges on their diffusion behavior as two receptors, or receptor nanoclusters, co-diffuse. In the first case, the nano-environment constitutes the intra-nanocluster milieu, whereas, in the second case, it corresponds to the inter-nanocluster surrounding. Surprisingly, only a modest percentage (< 30%) of all the recorded trajectories exhibited cross-correlation amplitudes above the background. These results are at first sight unexpected, considering that a large majority of the DC-SIGN receptors partition in nanoclusters[42], and thus a high level of receptor co-diffusion was anticipated.

To get more insight into these intriguing results and the nature of the broad range of $\tau_{x,1/2}$ values obtained from FCCS curves, we performed burst analysis on individual DC-SIGN trajectories. We first measured the time duration of individual bursts and calculated the diffusion coefficient ($D$) from individual bursts (Figure 4A). A broad range of $D$ values was obtained, as expected for most transmembrane receptors[43,48]. However, in contrast to our earlier results obtained using SPT[43], the $D$ values obtained here are significantly higher (peak ~0.5 μm²/s here *vs.* ~0.1 μm²/s from SPT measurements[43]). Moreover, while the $D$ distribution from SPT measurements showed a long tail towards values < 0.1 μm²/s[43], the $D$ distribution obtained from burst analysis at the nanoscale shows a tail towards $D$ values > 0.5 μm²/s. Since our FCS experiments are performed with much higher temporal resolution (~3μs) as compared to SPT (typically 20-50 ms), the $D$ values reported here most likely correspond to the diffusion of individual DC-SIGN molecules rather than to nanocluster diffusion as reported in Ref[43]. With most DC-SIGN receptors contained within nanoclusters[42], these results, therefore, imply that DC-SIGN can diffuse *inside* nanoclusters. Earlier experiments by Jacobson and co-workers indeed suggested that DC-SIGN nanoclusters are not fully packed and that certain lipids could freely diffuse within these nanoclusters[45].

To further support these observations, we measured the distribution of lag times between the end and the onset of consecutive bursts, $\tau_{off}$. We reasoned that if DC-SIGN is mostly organized within nanoclusters, its relative molecular density would be much larger inside nanoclusters than outside, which should be reflected in two markedly different inter-burst timescales. In agreement with the hypothesis, the histograms of $\tau_{off}$ show a clear bimodal distribution (Figure 4B) with one population with peaking at ~5 ·10$^{-3}$ s and a second one at ~2 s, regardless of the label used. We interpret these results based on the fact that the illumination area provided by the BNA (~50 nm) is much smaller than the reported average cluster size for DC-SIGN (~180 nm)[43]. Thus, when a nanocluster is present below the BNA, individual molecules inside the nanocluster will diffuse fast through the BNA gap, with short off-intervals between them, giving rise to the population of small $\tau_{off}$ values. On the other hand, the individual nanoclusters have a lower density on the membrane and diffuse much slower, so it will take a much longer time for a new nanocluster to arrive at the BNA illumination region, resulting in a population with longer $\tau_{off}$ values (Figure 4C). The random diffusion of individual receptors within large nanoclusters also explains the low occurrence of cross-correlation curves obtained in our measurements. To the best of our knowledge, these experiments resolve, for the first time, molecular diffusion inside receptor nanoclusters and distinguish it from the diffusion of the nanoclusters themselves.

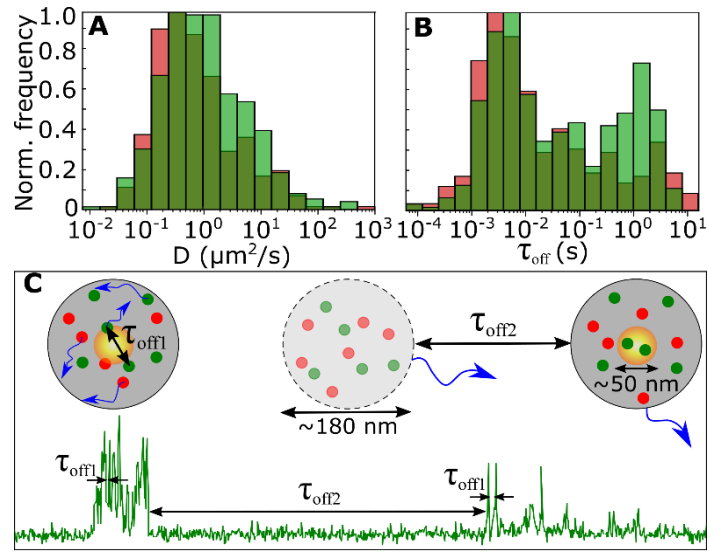

**Figure 4: Burst analysis reveals molecular diffusion of receptors inside nanoclusters as well as diffusion of nanoclusters on the plasma membrane of living cells. (A)** Histogram of the diffusion coefficients of DC-SIGN calculated from the burst length duration for multiple Atto520 (green, 454 bursts) and Atto647N (red, 364 bursts) trajectories. **(B)** Distribution of $\tau_{off}$ for the same trajectories as analyzed in **(a)**. **(C)** Schematic depiction of molecular diffusion inside nanoclusters (left) that give rise to the short $\tau_{off}$ timescales and the nanocluster diffusion (right) that lead to the long $\tau_{off}$ timescale as probed by the BNA. The grey circle illustrates a nanocluster containing multiple DC-SIGN molecules (green and red dots) that diffuse within the illumination area provided by the BNA gap (yellow circle). An experimentally obtained trajectory is shown below (1 ms binning, ~1 s length), denoting the two different $\tau_{off}$ regimes.

In summary, we have demonstrated that probe-based broadband BNA photonic antennas are highly effective in obtaining fast dynamics of membrane components, i.e., lipids and proteins, at the nanoscale in living cells. This is both owing to their large optical throughput and spatial confinement of the excitation field. In addition, we showed that BNAs provide comparable enhancement and spatial confinement to dimensions as small as ~50 nm for different excitation wavelengths. We demonstrated FCCS at the nanoscale in living cells and exploited the broadband behavior of these photonic antennas to investigate nanoscale co-diffusion of membrane receptors in living cells. Importantly, because of the nanometric dimensions of the BNA gap sizes and the sizes of the nanoclusters under study, we reveal for the first time intermolecular diffusion inside nanoclusters and receptor nanocluster diffusion. Altogether, our work shows that broadband photonic antennas are promising candidates for quantitative multi-color studies in living cell membranes with ultra-high spatiotemporal resolution.

## MATERIALS AND METHODS

***Antenna probe fabrication.*** The BNA probes were fabricated as described previously[36]. Briefly, heat-pulled optical fibers were coated with 5 nm Ti and 150 nm Al. The coated fibers were milled by focused ion-beam (FIB) to obtain a 500-700 nm diameter opening. This end-facet was subsequently coated with a high-quality Al layer of about 120 nm. Finally, the BNA was directly milled face-on into the Al-coated end-facet. This fabrication method allowed for extreme reproducibility of BNA probes with gap regions around 50 nm[36].

***Optical setup.*** For excitation in our combined NSOM/confocal setup, we used two lasers: a He-Ne laser at $\lambda = 633$ nm and an argon-krypton laser (Model 3060; Spectra-Physics, Santa Clara, CA) at $\lambda = 488$ nm. A combination of a polarizer and a $\lambda/2$ waveplate for each separate excitation wavelength ensured control of the incoming polarization to the antenna. The fluorescence from single molecules diffusing through the excitation volume of the antenna was collected by a high NA objective (oil, 1.3NA), split into two branches using a dichroic mirror (600 LP), individually filtered from the excitation light (Semrock 536/40 and 675/67) and detected by two single photon counting avalanche photodiodes (APDs) A photon-counting unit (NI BCN-800) was used to record photon arrival time traces that are successively processed by a software correlator. A shear-force feedback system on the antenna probe, together with the piezo-electric sample stage, guaranteed close and constant distance regulation between the antenna and the cell membrane. The feedback system is based on a piezo-electric tuning fork that is operated in the air while the end facet of the probe containing the BNA is immersed in liquid using the diving bell concept[37].

***FDTD simulations.*** 3D numerical modeling of the antenna was based on finite-difference time-domain (FDTD) simulations. The simulations consider a volume spanning ±2.6 μm in *x* and *y* around the BNA end face. The refraction index and taper angle of the dielectric body of the probe were chosen to be 1.448 and 32°, respectively, and the aluminum dielectric constant is given by the Drude model adapted for each wavelength considered. The BNA is located at x = y = z = 0. In the z direction, the simulation extends to 1 μm in air and terminates at 7 μm into the body of the probe. All six boundaries of the computation volume are terminated with convolutional-periodic matching layers to avoid parasitic unphysical reflections around the probe. The non-uniform grid resolution varies from 25 nm for portions at the periphery of the simulation to 5 nm for the region in the immediate vicinity of the BNA (±200 nm in x and y and −200 to 100 nm in z). Excitation was done by a linearly polarized Gaussian beam launched at 7 μm away from the tip body and propagating towards the BNA.

***Cells and samples preparation.*** Chinese hamster ovary (CHO) cells were cultured in phenol-red free Dulbecco's Modified Eagle Medium (DMEM) with nutrient mixture F-12 (1:1) supplemented with 10% fetal calf serum and Antibiotic Antimycotic Solution (Gibco). For incorporation of Atto647N conjugated phosphoethanolamine lipid (PE) in the cell membrane, we followed previously published protocols[16,19]. Bovine Serum Albumin (BSA)/PE analog complexes were prepared by dissolving the lipid analogs in $CHCl_3$/MeOH (3:1). From the stock solution, 100 nM of lipid analog were dried under a stream of nitrogen and re-dissolved in 20 μl of absolute ethanol. After adding 1 ml of defatted BSA (0.2 mM in DMEM), the solution was vigorously vortexed. Cells spontaneously adhered to a glass coverslip after 48 hours of incubation at 37° C. For PE analogs incorporation, cells were washed with DMEM and incubated with BSA/PE analog complexes for 10 min at 25° C temperature, washed with DMEM, and prepared for observation. Typical concentrations of BSA/PE complexes were 100–300 nM.

For the experiments on DC-SIGN, CHO cell lines stably expressing DC-SIGN wild type, containing a short C-terminal AU1 tag as already published[43], were cultured in Ham's F-12 medium (LabClinics) supplemented with 10% fetal calf serum and Antibiotic Antimycotic Solution (Gibco). Monovalent single-chain anti-human AU1 antibodies (mAbs) were generated from AU1 Ab (Covance) by reduction with Dithiothreitol (DTT, Invitrogen) according to manufacturer´s instructions. Reduced Abs were then labeled with either Atto520 or Atto647N according to standard protocols provided by the manufacturer. Glass coverslip-adhered CHO cells were incubated for 30 min at room temperature using equimolar concentrations of Atto520- and Atto647N-conjugated single chain mAbs at saturating conditions to label all DC-SIGN receptors. Before imaging, extensive washing with the serum-free medium was performed to remove non-bound mAbs.

***Burst analysis.*** Fluorescence bursts were detected and quantified from unfiltered photon arrival-time recordings as previously described[39]. In brief, a likelihood-based algorithm was used to sequentially analyze photon recordings to test the null hypothesis (no burst, recording compatible with background noise) against the hypothesis that a fluorescence burst arises as a consequence of a fluorophore crossing the excitation volume. Background level and typical fluorophore intensity were estimated from the trace to be analyzed. Probabilities associated with false positives and missing event errors were both set to 0.001[32,49]. From all detected bursts, intensity and length are measured, as well as the time between consecutive bursts for the DC-SIGN experiments.

**Supporting Information**

Supporting Information is available from the Wiley Online Library or from the author.

**Acknowledgements**

The research leading to these results has received funding from the European Commission H2020 Program under grant agreement ERC Adv788546 (NANO-MEMEC), Government of Spain (Severo Ochoa CEX2019-000910-S, State Research Agency (AEI) PID2020-113068RB-I00 / 10.13039/501100011033 (to M.F.G.-P., BES-2015-072189 (to M.S.-P.), grant RYC-2015-17896 funded by MCIN/AEI/10.13039/501100011033 and "El FSE invierte en tu futuro" (to C.M.), grants BFU2017-85693-R and PID2021-125386NB-I00 funded by MCIN/AEI/10.13039/501100011033/ and FEDER "Una manera de hacer Europa" (to C.M.), Fundació CELLEX (Barcelona), Fundació Mir-Puig and the Generalitat de Catalunya through the CERCA program and AGAUR (Grant No. 2017SGR1000 to M.F.G.-P. and 2017SGR940 to C.M.).

**Conflict of Interest**

The authors declare no conflict of interest.

**Data Availability Statement**

The data that support the findings of this study are available from the corresponding author upon reasonable request.

## REFERENCES

[1]  K. Simons, D. Toomre, *Nat Rev Mol Cell Biol* **2000**, *1*, 31.

[2]  D. Lingwood, K. Simons, *Science* **2010**, *327*, 46.

[3]  D. A. Brown, E. London, *Annu. Rev. Cell Dev. Biol.* **1998**, *14*, 111.

[4]  A. Kusumi, C. Nakada, K. Ritchie, K. Murase, K. Suzuki, H. Murakoshi, R. S. Kasai, J. Kondo, T. Fujiwara, *Annu. Rev. Biophys. Biomol. Struct.* **2005**, *34*, 351.

[5]  M. F. Garcia-Parajo, A. Cambi, J. A. Torreno-Pina, N. Thompson, K. Jacobson, *J. Cell Sci.* **2014**, *127*, 4995.

[6]  A. Kusumi, T. K. Fujiwara, N. Morone, K. J. Yoshida, R. Chadda, M. Xie, R. S. Kasai, K. G. N. Suzuki, *Semin. Cell Dev. Biol.* **2012**, *23*, 126.

[7]  K. Gowrishankar, S. Ghosh, S. Saha, C. Rumamol, S. Mayor, M. Rao, *Cell* **2012**, *149*, 1353.

[8]  E. Sezgin, I. Levental, S. Mayor, C. Eggeling, *Nat. Publ. Gr.* **2017**, DOI 10.1038/nrm.2017.16.

[9]  P. Sil, N. Mateos, S. Nath, S. Buschow, C. Manzo, K. G. N. Suzuki, T. Fujiwara, A. Kusumi, M. F. Garcia-Parajo, S. Mayor, **n.d.**, DOI 10.1091/mbc.E18-11-0715.

[10] C. Eggeling, C. Ringemann, R. Medda, G. Schwarzmann, K. Sandhoff, S. Polyakova, V. N. Belov, B. Hein, C. Von Middendorff, A. Schönle, S. W. Hell, *Nature* **2009**, *457*, DOI 10.1038/nature07596.

[11] P. M. Winkler, M. F. García-Parajo, *Biochem. Soc. Trans.* **2021**, *49*, 2357.

[12] G. Vicidomini, H. Ta, A. Honigmann, V. Mueller, M. P. Clausen, D. Waithe, S. Galiani, E. Sezgin, A. Diaspro, S. W. Hell, C. Eggeling, *Nano Lett* **2015**, *15*, 29.

[13] R. Schmidt, T. Weihs, C. A. Wurm, I. Jansen, J. Rehman, S. J. Sahl, S. W. Hell, **n.d.**, DOI 10.1038/s41467-021-21652-z.

[14] Y. Lee, C. Phelps, T. Huang, B. Mostofian, L. Wu, Y. Zhang, K. Tao, Y. H. Chang, P. J. S. Stork, J. W. Gray, D. M. Zuckerman, X. Nan, *Elife* **2019**, *8*, DOI 10.7554/ELIFE.46393.

[15]  C. Manzo, T. S. Van Zanten, M. F. Garcia-Parajo, *Biophys. J.* **2011**, *100*, L8.

[16]  C. Eggeling, C. Ringemann, R. Medda, G. Schwarzmann, K. Sandhoff, S. Polyakova, V. N. Belov, B. Hein, C. von Middendorff, A. Schönle, S. W. Hell, *Nature* **2009**, *457*, 1159.

[17]  M. J. Levene, J. Korlach, S. W. Turner, M. Foquet, H. G. Craighead, W. W. Webb, *Science* **2003**, *299*, 682.

[18]  C. V. Kelly, D. L. Wakefield, D. A. Holowka, H. G. Craighead, B. A. Baird, *ACS Nano* **2014**, *8*, 7392.

[19]  C. Manzo, T. S. Van Zanten, M. F. Garcia-Parajo, *Biophys. J.* **2011**, *100*, L8.

[20]  R. Strack, *Nat. Methods* **2015**, *12*, 1111.

[21]  G. Donnert, J. Keller, C. A. Wurm, S. O. Rizzoli, V. Westphal, A. Schönle, R. Jahn, S. Jakobs, C. Eggeling, S. W. Hell, *Biophys. J.* **2007**, *92*, 67.

[22]  J. Wenger, D. Gérard, P.-F. Lenne, H. Rigneault, J. Dintinger, T. W. Ebbesen, A. Boned, F. Conchonaud, D. Marguet, *Opt. Express* **2006**, *14*, 12206.

[23]  D. Punj, M. Mivelle, S. B. Moparthi, T. S. van Zanten, H. Rigneault, N. F. van Hulst, M. F. García-Parajó, J. Wenger, *Nat. Nanotechnol.* **2013**, *8*, 512.

[24]  A. Puchkova, C. Vietz, E. Pibiri, B. Wu, M. Sanz Paz, G. P. Acuna, P. Tinnefeld, *Nano Lett.* **2015**, *15*, 8354.

[25]  T. S. van Zanten, M. J. Lopez-Bosque, M. F. Garcia-Parajo, *Small* **2010**, *6*, 270.

[26]  T. Lohmüller, L. Iversen, M. Schmidt, C. Rhodes, H. L. Tu, W. C. Lin, J. T. Groves, *Nano Lett.* **2012**, *12*, 1717.

[27]  V. Flauraud, T. S. Van Zanten, M. Mivelle, C. Manzo, M. F. Garcia Parajo, J. Brugger, *Nano Lett.* **2015**, *15*, 4176.

[28]  B. Pradhan, S. Khatua, A. Gupta, T. Aartsma, G. Canters, M. Orrit, *J. Phys. Chem. C* **2016**, *120*, 25996.

[29]  J. D. Flynn, B. L. Haas, J. S. Biteen, **2015**, DOI 10.1021/acs.jpcc.5b08049.

[30]  S. A. Lee, J. S. Biteen, *J. Phys. Chem. C* **2018**, *122*, 25.

[31] A. Krasnok, M. Caldarola, N. Bonod, A. Alú, *Adv. Opt. Mater.* **2018**, *6*, 1.

[32] R. Regmi, P. M. Winkler, V. Flauraud, K. J. E. Borgman, C. Manzo, J. Brugger, H. Rigneault, J. Wenger, M. F. García-Parajo, *Nano Lett.* **2017**, *17*, 6295.

[33] P. M. Winkler, R. Regmi, V. Flauraud, J. Brugger, H. Rigneault, J. Wenger, M. F. García-Parajo, *J. Phys. Chem. Lett.* **2018**, *9*, 110.

[34] P. M. Winkler, R. Regmi, V. Flauraud, J. Brugger, H. Rigneault, J. Wenger, M. F. García-Parajo, *ACS Nano* **2017**, *11*, 7241.

[35] D. Magatti, F. Ferri, *Appl. Opt.* **2001**, *40*, 4011.

[36] M. Mivelle, T. S. Van Zanten, L. Neumann, N. F. Van Hulst, M. F. Garcia-Parajo, *Nano Lett.* **2012**, *12*, 5972.

[37] M. Koopman, B. I. De Bakker, M. F. Garcia-Parajo, N. F. Van Hulst, *Appl. Phys. Lett.* **2003**, *83*, 5083.

[38] I. A. Ibrahim, M. Mivelle, T. Grosjean, J.-T. Allegre, G. W. Burr, F. I. Baida, *Opt. Lett.* **2010**, *35*, 2448.

[39] K. Zhang, H. Yang, *J. Phys. Chem. B* **2005**, *109*, 21930.

[40] R. Guo, E. C. Kinzel, Y. Li, S. M. Uppuluri, A. Raman, X. Xu, *Opt. Express* **2010**, *18*, 4961.

[41] G. Lu, W. Li, T. Zhang, S. Yue, J. Liu, L. Hou, Z. Li, Q. Gong, *ACS Nano* **2012**, *6*, 1438.

[42] B. I. De Bakker, F. De Lange, A. Cambi, J. P. Korterik, E. M. H. P. Van Dijk, N. F. Van Hulst, C. G. Figdor, M. F. Garcia-Parajo, *ChemPhysChem* **2007**, *8*, 1473.

[43] C. Manzo, J. A. Torreno-Pina, B. Joosten, I. Reinieren-Beeren, E. J. Gualda, P. Loza-Alvarez, C. G. Figdor, M. F. Garcia-Parajo, A. Cambi, *J. Biol. Chem.* **2012**, *287*, 38946.

[44] J. A. Torreno-Pina, B. M. Castro, C. Manzo, S. I. Buschow, A. Cambi, M. F. Garcia-Parajo, *Proc. Natl. Acad. Sci. U. S. A.* **2014**, *111*, 11037.

[45] M. S. Itano, A. K. Neumann, P. Liu, F. Zhang, E. Gratton, W. J. Parak, N. L. Thompson, K. Jacobson, *Biophys. J.* **2011**, *100*, 2662.

[46] A. Cambi, F. De Lange, N. M. Van Maarseveen, M. Nijhuis, B. Joosten, E. M. H. P. Van Dijk, B. I. De Bakker, J. A. M. Fransen, P. H. M. Bovee-Geurts, F. N. Van Leeuwen, N. F. Van Hulst, C. G. Figdor, *J. Cell Biol.* **2004**, *164*, 145.

[47] P. Liu, X. Wang, M. S. Itano, A. K. Neumann, K. Jacobson, N. L. Thompson, *Traffic* **2012**, *13*, 715.

[48] T. S. van Zanten, A. Cambi, M. Koopman, B. Joosten, C. G. Figdor, M. F. Garcia-Parajo, *PNAS* **2009**, *106*, 18557.

[49] V. Flauraud, T. S. Van Zanten, M. Mivelle, C. Manzo, M. F. G. Parajo, J. Brugger, *Nano Lett.* **2015**, *15*, 4176.

Supporting Information

# Broadband plasmonic nanoantennas for multi-color nanoscale dynamics in living cells

*Maria Sanz-Paz[†,¥,#], Thomas S. van Zanten[†,‡,#], Carlo Manzo[†,£], Mathieu Mivelle[§,\*], Maria F. Garcia-Parajo[†,⊥,\*]*

[†] ICFO-Institut de Ciencies Fotoniques, The Barcelona Institute for Science and Technology, 08860 Barcelona, Spain;

[¥] Department of Physics, University of Fribourg, Chemin du Musée 3, Fribourg CH-1700, Switzerland; [‡] National Centre for Biological Sciences, Bangalore, India;

[£] Facultat de Ciéncies, Tecnologia i Enginyeries, Universitat de Vic – Universitat Central de Catalunya, C. de la Laura 13, 08500 Vic, Spain;

[§] Sorbonne Université, CNRS, Institut des NanoSciences de Paris, UMR 7588, 75005 Paris, France;

[⊥] ICREA, Pg. Lluis Companys 23, 08010 Barcelona, Spain

[#] Equally contributing authors

## Corresponding Authors

*Email: maria.garcia-parajo@icfo.eu

*Email: mathieu.mivelle@sorbonne-universite.fr

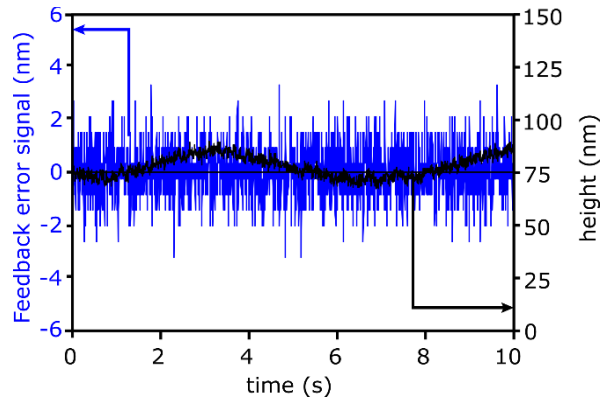

**Figure S1:** Axial fluctuations of the cell membrane as a function of time as the antenna probe is maintained at a constant distance of about 10 nm above the cell (right vertical axis). The left vertical axis corresponds to the error signal of the feedback loop used to keep the membrane-antenna distance constant. The right axis corresponds to the height of the tip with respect to the substrate. The fluctuations of the tip observed due to the movement of the cell membrane are of the order of 25 nm.

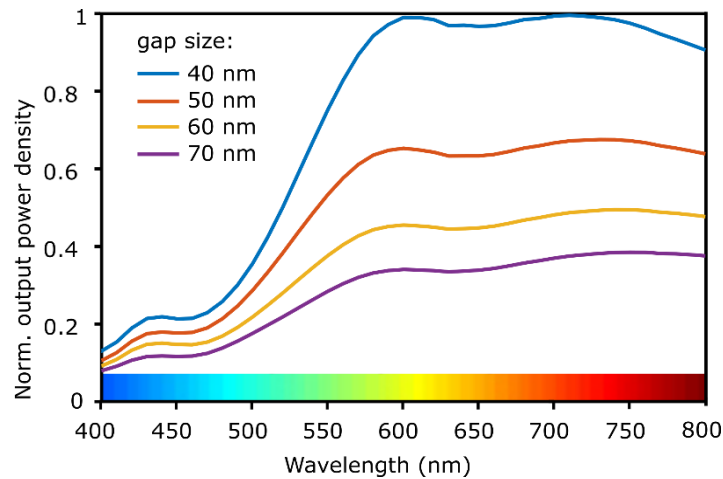

**Figure S2:** FDTD simulations of the broadband BNA response in the visible for different gap sizes. The incoming optical field is transversally polarized to the antenna gap so that excitation is optimal. Changing the gap size does not affect the spectral response, but the enhancement increases with decreasing gap size.

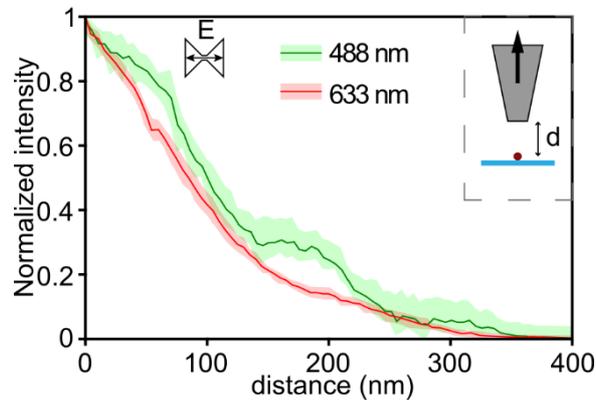

**Figure S3:** Normalized power density as a function of axial distance from the antenna to the beads, for both excitation wavelengths and longitudinal polarization. A single exponential fitting of the yields a field penetration (1/e) of 108.9±1.0 nm at 633 nm, and of 141.9±3.1 nm at 488 nm.

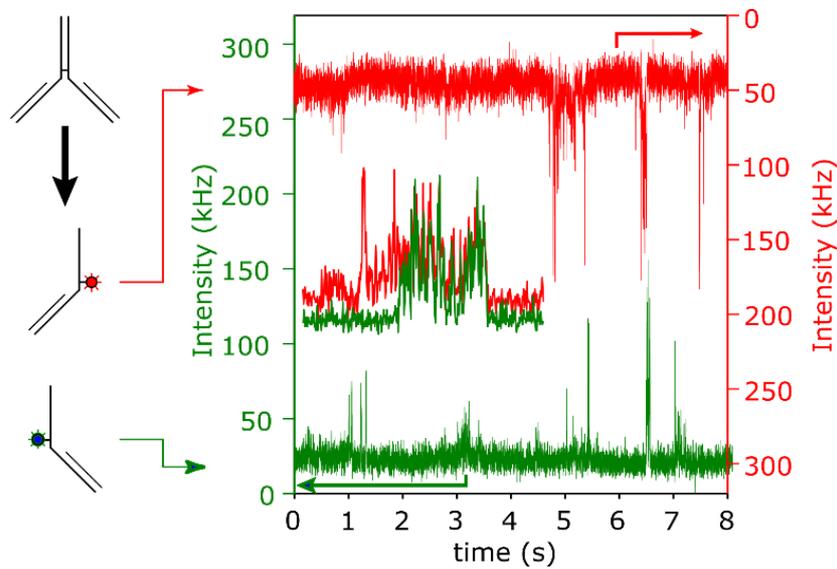

**Figure S4:** Representative fluorescent time traces (1 ms bin) of Atto520 (green) and Atto647N (red) conjugated to single chain antibodies and bound to DC-SIGN, expressed on CHO cells. Trajectories were generated using transversally polarized excitation of a BNA probe simultaneously excited with λ = 488nm and λ = 633 nm. The scheme on the left shows the labeling strategy.

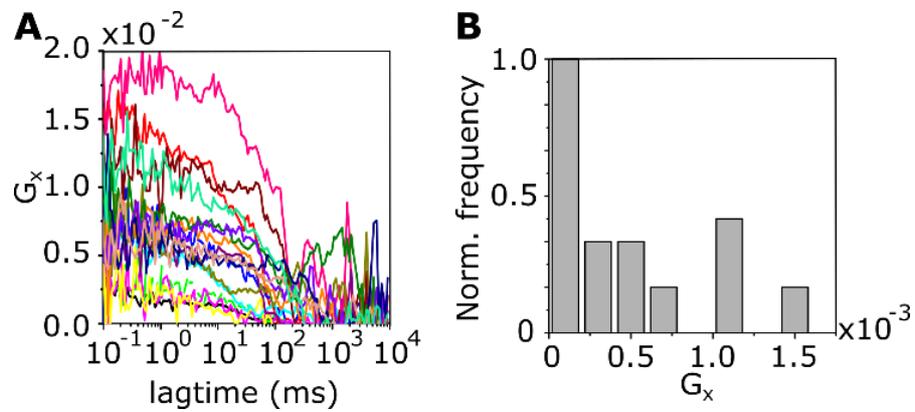

**Figure S5: Statistics on the nanoscale cross-correlation of dual-color labeled DC-SIGN. (A)** CCF of 17 different dual-color time traces indicating positive DC-SIGN cross-correlation, obtained from different cells and/or regions of the cell membrane. **(B)** Amplitudes for the CCF curves where no cross-correlation between DC-SIGNs was observed.